\documentclass[12pt]{iopart}
\usepackage[utf8]{inputenc}
\usepackage[T1]{fontenc}
\usepackage{lmodern} %
\usepackage{iopams} 
\usepackage{bm}
\usepackage{graphicx}
\usepackage{hyperref}

\newcommand{\pd}[2]{\frac{\partial{#1}}{\partial{#2}}}

\newcommand{\bra}[1]{\langle#1|}
\newcommand{\ket}[1]{|#1\rangle}
\newcommand{\proj}[2]{| #1 \rangle \! \langle #2 |}
\newcommand{\braket}[1]{\langle #1 \rangle}

\begin{document}

\title{Optimal number of pigments in photosynthetic complexes}

\author{Simon Jesenko and Marko \v Znidari\v c}

\address{Faculty of Mathematics and Physics, University of Ljubljana, Slovenia}
\begin{abstract}
We study excitation energy transfer in a simple model of photosynthetic complex. The model, described by Lindblad equation, consists of pigments interacting via dipole-dipole interaction. Overlapping of pigments induces an on-site energy disorder, providing a mechanism for blocking the excitation transfer. Based on the average efficiency as well as robustness of random configurations of pigments, we calculate the optimal number of pigments that should be enclosed in a pigment-protein complex of a given size. The results suggest that a large fraction of pigment configurations are efficient as well as robust if the number of pigments is properly chosen. We compare optimal results of the model to the structure of pigment-protein complexes as found in nature, finding good agreement.
\end{abstract}


\section{Introduction}

Photosynthesis is the main natural process for harvesting Sun's energy on Earth, providing a food source for a great variety of organisms ranging from highly evolved photosynthetic systems in higher plants to simpler bacteria \cite{BlankenshipBook}. In spite of such diversity, basic underlying principles are shared among majority of light harvesting organisms. The initial stage of photosynthesis usually involves multiple \textit{pigment protein complexes} (PPC) that consist of a number of pigment molecules (i.e., chromophores) held in place by a protein cage. PPCs are employed to absorb incoming photons as well as to transport the resulting excitation to the \textit{reaction center}, where the excitation is used to initiate chemical reactions. The absorption of light and in particular channeling of the absorbed energy to the reaction center is known to achieve high efficiency \cite{BlankenshipBook}.

One of the most studied PPCs is the Fenna-Matthews-Olson (FMO) complex that is found in the photosynthetic apparatus of \textit{green sulfur bacteria}. It is the first PPC for which the atomic structure has been determined~\cite{matthews1979structure}. FMO is composed of a large protein envelope that encloses a tightly packed group of 7 pigment molecules\footnote{Recent structural analysis \cite{Tronrud2009} suggests also the presence of an eighth BChl that is weakly bound to each monomeric unit as an additional input site. Its distance from the core 7 pigments is quite large and therefore we do not take it into account in our study.} called bacteriochlorophylls (BChl). In FMO the main role of BChl pigments is actually not to absorb light but instead to transport electronic excitations from the input pigment, which is close to the antenna complex, towards a ``sink'' pigment channeling excitation to the reaction center. Recently discovered long-lasting quantum coherence in FMO complex \cite{Engel2007} gave an additional boost to studies of excitation transport in PPCs. Previously it was namely believed that the transport in PPCs is predominantly of a classical nature. Many aspects of time dependence have been studied \cite{Brixner2005,Cho2005, Ishizaki2009}, including the functional role of quantum coherences  \cite{Rebentrost2009a, Ishizaki2010PCCP, Pachon2011, Ishizaki2009}, as well as the possibility of transport enhancement via environmental interaction \cite{Plenio2008, Rebentrost2009,Mohseni2008,Caruso2009}.

Also important is the question of structural characteristics of PPCs that enable efficient excitation transfer, for instance, why has a given complex precisely the shape found in nature. Positions and orientations of pigments, known with high precision from crystallographic measurements, on casual inspection show no clear ordering or organization that would enable an easy classification of efficient configurations. It is also not clear, how \textit{special} efficient configurations are - whether efficient configurations are a result of a long-lasting process of evolutionary improvement, or, can efficient configurations be readily achieved probing few random configurations of pigments. Prior to the availability of crystallographic structural data, the average minimal distance between pigment molecules was estimated based on the comparison between florescence yield of \textit{in vivo} and \textit{in vitro} chlorophyll solutions \cite{beddard1976concentration}. Recently, the efficiency of random configurations within simplified models of PPC was inspected \cite{Scholak2011, Scholak2011a, Scholak2011b, Mohseni2011}, suggesting that the efficient configurations are relatively probable. This complies with the results obtained for the Photosystem I from plants and cyanobacteria \cite{Sener2002, Sener2005a}, where random orientations of pigment molecules were probed, and high efficiencies of excitation energy transfer (EET) were obtained irrespectively of the pigment orientation. Also, the PPC configuration was shown to be robust to the removal of a pigment from the complex \cite{Sener2002}.

Fundamental, yet still unanswered question that we address in present work is, why has a particular photosynthetic complex exactly the specified number of pigments. In other words, what is the optimal number of pigments for a given size of the complex\footnote{Note that the size of the PPC might be fixed by external factors like for instance the membrane thickness.}, or, equivalently, what is the optimal size of the complex for a given number of pigments. For instance, why do we find exactly 7 pigments in the FMO complex and not more or less. Using a simple model whose parameters are taken from experiments, that is without any fitting parameters, we calculate the optimal number of photosynthetic pigment molecules for different complex sizes and compare these theoretical predictions with the actual number of pigments found in naturally occurring complexes. We find a very good agreement for a variety of PPCs in different organisms. To judge the optimality we use two criteria: (i) the average efficiency of the excitation transfer, where the averaging is performed over random configurations of pigment molecules, and (ii) the robustness of the efficiency to small variations of pigment's locations. A rationale behind these two choices is that ``good'' PPCs should have high efficiency but at the same time also be robust. A specially ``tuned'' configuration, that has a very high efficiency which though is very fragile, will obviously not work in a natural environment with its changing conditions. Additionally, from an evolutionary perspective, the efficiency should be stable with respect to different foldings of the protein cage. It is advantageous to have a PPC with such a number of pigments that will results in high average efficiency, i.e., in many close-to-optimal pigment configurations. Thereby, a small change in the environment, be it of a chemical origin or for instance a genetic mutation, will still result in a functional PPC. Robustness of quantum coherence to structural changes in the PPCs has been also found experimentally~\cite{Hayes2011}. The model we use to describe the excitation transfer across the PPC consists of a Lindblad master equation describing a dipole-coupled pigments with an on-site excitonic energies being determined by the distances between disc-like pigments. If two discs come too close, i.e., if they overlap, this effectively rescales their energies, introducing a disorder. We should say that the optimal number of pigments that we predict is quite insensitive to details of the underlying model. Agreement between predictions of our model and naturally-occurring PPCs shows that Nature has optimized PPCs by using just the right number of pigments so that the resulting PPCs are highly efficient and robust at the same time.

\section{The model}

To calculate the efficiency of the excitation transfer in the PPC, we need equations of motion describing dynamics of excitation on multiple chromophores, coupled to the environment. It is the ratio of chromophore-chromophore interaction strength to the chromophore-environment coupling that determines the applicability of various models. When interaction between chromophores is small compared to the environmental coupling the F\"orster theory \cite{forster1948} is applicable, leading to a picture of incoherent hopping of excitation between chromophores. In the opposite limit of strong inter-chromophore interaction and weak coupling to the environment, the excitation dynamics can be described by quantum master equation, either the Redfield equation or, employing a secular approximation, the Lindblad equation \cite{Ishizaki2009b}. For nonperturbative parameter ranges, more advanced methods have been developed \cite{Ishizaki2009a, Nalbach2011,Ritschel2011}, usually at the expense of higher computational complexity.

Our main optimality criterion, namely the \textit{transfer efficiency}, is relatively insensitive to details of excitation time-dependence. Thus, we are going to use the simplest description of EET with the Lindblad equation, which retains coherent nature of transport, while still taking environmental interactions into account. We expect that more exact descriptions, which in general enhance excitonic oscillations, lead to similar results due to our averaging procedure. Also, these oscillations appear on the time scales of few $100\,{\rm fs}$, which is much shorter than the time scale of excitonic transfer.

In the following, we will introduce the Lindblad equation for the \textit{overlapping disc model}, used for the description of excitation dynamics in PPCs. Optimality criteria for the efficiency of PPC configurations as used in latter sections will also be presented.

\subsection{Lindblad master equation}

Internal dynamics of the system of $N$ chromophores within a single-excitation manifold is determined by the Hamiltonian of the form \cite{may2011charge}
\begin{equation}
  \label{eq:3}
  H = \sum_{n=1}^N \epsilon_n \proj{n}{n} + \sum_{n \neq m = 1}^N V_{mn} (\proj{m}{n} + \proj{n}{m}),
\end{equation}
where a state $\ket{n}$ represents an excitation on the $n$-th chromophore site, i.e., the electronic state of the $n$-th chromophore being in the 1st excited state. Because EET is sufficiently fast, events with two excitations being present at the same time are rare and it is sufficient to consider only zero and single-excitation subspace~\cite{Ishizaki2010PCCP}. The coupling $V_{mn}$ is due to dipole-dipole interaction between chromophores of the form
\begin{equation}
  \label{eq:4}
  V_{mn} = \frac{1}{4 \pi \varepsilon_0}\left(\frac{\bm d_m\cdotp \bm d_n}{r_{mn}^3}-3 \frac{(\bm d_m \cdotp \bm r_{mn})(\bm d_n \cdot \bm r_{mn})}{r_{mn}^5}\right),
\end{equation}
where $\bm r_{mn}=\bm x_m - \bm x_n$ is a vector, connecting the $m$-th and $n$-th chromophores, $\bm d_n$ is a transition dipole moment between the ground and the 1st excited state of the $n$-th chromophore.

Because the system of chromophores is coupled to the protein and nuclear degrees of freedom it is described by a reduced density matrix $\rho$. Decoherence due to environmental interaction, recombination of excitation to the ground state, and transfer of excitation to the sink, are modeled by Lindblad superoperators that augment the von Neumann equation for the time evolution of density matrix,
\begin{equation}
  \label{eq:5}
  \dot \rho = -i [H, \rho] + \mathcal{L}_{\rm{deph}}(\rho) + \mathcal{L}_{\rm{sink}}(\rho) + \mathcal{L}_{\rm{recomb}}(\rho).
\end{equation}
To model the effects of the environment, we have taken a simplified picture of purely dephasing Lindblad superoperators (i.e. Haken-Strobl model), which is believed to capture the basic environmental effects and was used in various previous studies~\cite{Cao2009,Plenio2008,Mohseni2008,Rebentrost2009}. For longer time, relevant for the efficiency of PPC, it has been shown that the description with the Lindblad equation accounts for the main features of the dynamics~\cite{Wu2010,Wu2011}. Dephasing Lindblad superoperator destroys a phase coherence of any coherent superposition of excitations at different chromophores, and is given by
\begin{equation}
  \label{eq:6}
  \mathcal{L}_{\rm{deph}}(\rho)=2\gamma \sum_{n=1}^N \left( \proj{n}{n}\, \rho\, \proj{n}{n} - \frac{1}{2}\{ \proj{n}{n}, \rho \} \right),
\end{equation}
where a site-independent dephasing rate is given by $\gamma$ and $\{\,,\,\}$ represents the anticommutator. Irreversible transfer of excitation from the $N$-th chromophore to the sink $\ket{s}$ is modeled by Lindblad superoperator
\begin{equation}
  \label{eq:7}
  \mathcal{L}_{\rm{sink}}(\rho)=2\kappa \left(\proj{s}{N}\,\rho\,\proj{N}{s} - \frac{1}{2}\{\proj{N}{N},\rho \} \right),
\end{equation}
where $\kappa$ denotes the sink rate. The irreversible loss of excitation due to recombination is given by an analogous term
\begin{equation}
  \label{eq:8}
  \mathcal{L}_{\rm{recomb}}(\rho)=2\Gamma \sum_{n=1}^N \left(\proj{0}{n}\,\rho\,\proj{n}{0} - \frac{1}{2}\{\proj{n}{n},\rho \} \right),
\end{equation}
with a site-independent recombination rate given by $\Gamma$. The ground state of a chromophore system (state without any excitation) is represented as $\ket{0}$. Note that $\mathcal{L}_{\rm{sink}}$ and $\mathcal{L}_{\rm{recomb}}$ can be equivalently represented by an antihermitian Hamiltonian at the expense of non-conserved density matrix probability \cite{Rebentrost2009}, avoiding the need for an additional sink and the ground state.

Relevant environmental parameters going into Lindblad equation \eref{eq:5} are dephasing strength $\gamma$, recombination rate $\Gamma$ and the sink rate $\kappa$. We use standard values inferred from experiments and used before~\cite{Rebentrost2009}, sink rate $\kappa=1 \, \rm{ps^{-1}}$, recombination rate $\Gamma=1 \, \rm{ns^{-1}}$ and dephasing rate at room temperature $\gamma=300\,{\rm cm}^{-1}$\footnote{The values entering the Lindblad equation \eref{eq:5} should be in units of frequency, e.g. $\rm{s}^{-1}$. The conversion from inverse centimeters ${\rm cm}^{-1}$ as used traditionally in spectroscopy is given by $\omega [ {\rm s} ^{-1} ] = 200 \pi c \tilde{\nu}$, where $c = 3 \times 10^8$ and $\tilde \nu$ is value in $[{\rm cm^{-1}}]$.}. Dephasing rate, being a product of temperature and the derivative of the spectral density, can be estimated by using experimentally determined parameters of the spectral density (as done e.g. in reference~\cite{Rebentrost2009}). This value approximately agrees with the optimal dephasing rate at which transfer is most efficient~\cite{Rebentrost2009,Plenio2008}. We note that the results shown depend very weakly on the actual value of the dephasing rate as long as it is of the same order of magnitude as $\gamma=300\,{\rm cm}^{-1}$. In \ref{app:gamma} we show that the values of $\gamma=150\,{\rm cm}^{-1}$ and $600\,{\rm cm}^{-1}$ give almost the same optimal PPC size. 

\subsection{Overlapping discs model}

\begin{figure}[tb]
  \includegraphics{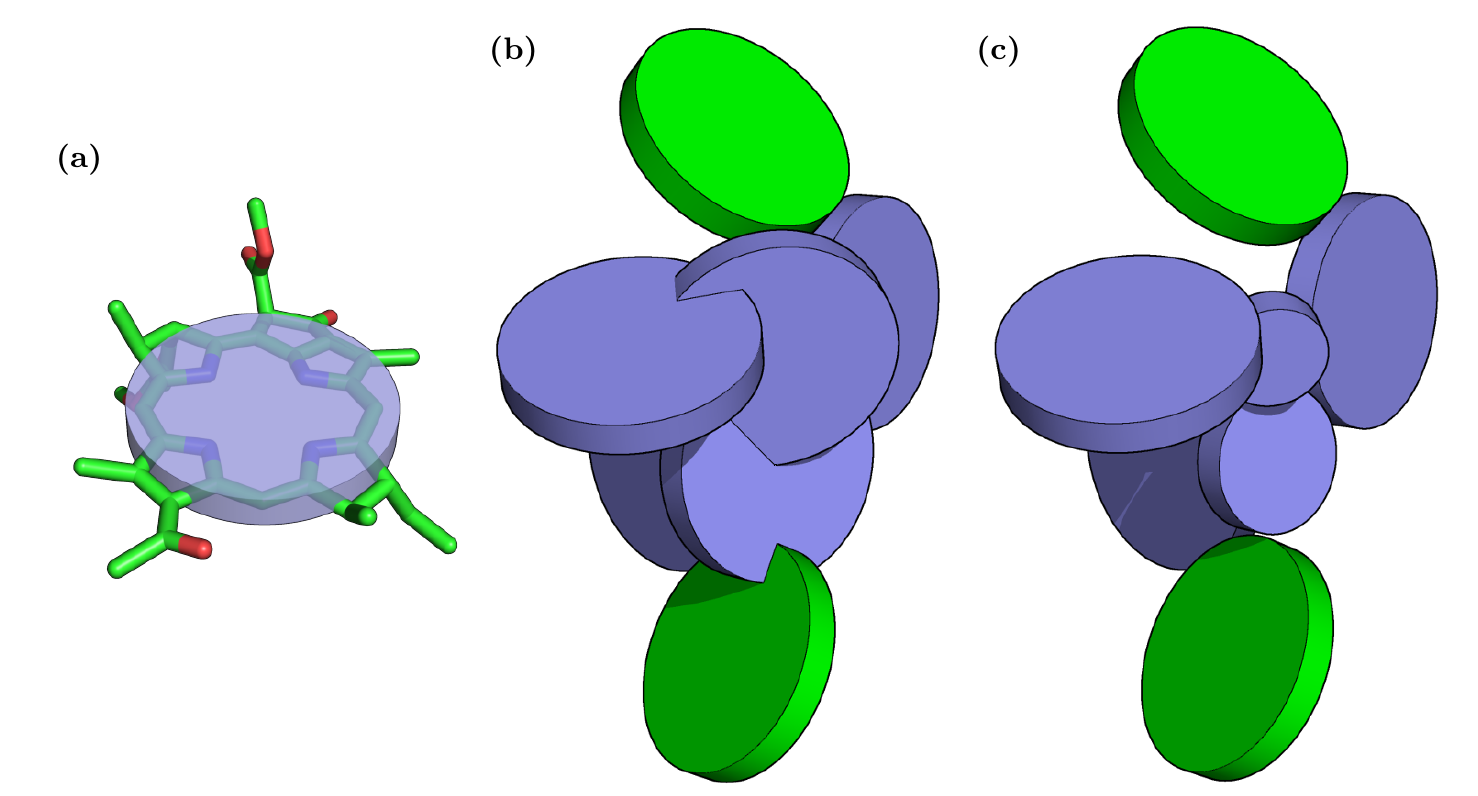}
  \caption{Illustration of the simple model of PPC. {\bf (a)} BChl molecule with omitted phytyl tail, enclosed by the cylinder as used in simulations ($a= 1\,\rm{\AA}$, $r=4 \, \rm{\AA}$). {\bf (b)} Discs, randomly positioned in sphere, prior to the reduction of radii of overlapping discs. Green discs at the top / bottom represent the input site / output site. {\bf (c)} Same sample as in (b) after the shrinking of overlapping discs. }
  \label{fig:ds_skica}
\end{figure}

Because we want to study the dependence of efficency on the size of PPC, keeping the number of pigments fixed, we have to account for the size-dependence of the Hamiltonian. On-site energies and interaction strengths in equation (\ref{eq:3}) will be determined from the geometry of pigments configuration. Changing the PPC's size two gross effects are at play. First, as the distance between pigments is reduced, the dipole-dipole interaction between pigments gets larger, enhancing the transfer of excitation among them; this effect is already taken into account by the $\sim 1/r^3$ dependence of $V_{nm}$ in equation (\ref{eq:4}). Secondly, as chromophores get even closer together, approaching the distances comparable to the extent of the chromophore electronic orbitals, the Hamiltonian \eref{eq:3} is not sufficient for the description of EET anymore because the effects due to electronic orbital overlap become important. A detailed analysis of processes that take place as chromophores get close together would require advanced quantum chemistry methods and is out of scope of this paper. However, the main effect can be effectively taken into account by appropriately rescaling parameters of equation \eref{eq:3}. Because the pigment molecules will be deformed, their excitation energies will also change. As the on-site energies $\epsilon_n$, being of the order of few ${\rm eV}$, are about $\sim 100$ times larger than $V_{nm}$, even a small relative change in $\epsilon_n$ can have a large effect. Effectively, close or even overlapping chromophores will therefore result in widely different values of on-site energies $\epsilon_n$ at different sites, i.e., in a disorder. Thus there are two competing factors that determine the optimal size of PPCs: reducing inter-chromophore distance increases EET, while the overlapping of chromophores introduces a disorder that effectively suppresses EET.

On-site energy has an additional contribution due to local environment (e.g. because of pigment-protein interactions), which is however of the order of $\sim 10^2\,{\rm cm}^{-1}$ and is usually much smaller that the on-site disorder due to pigment overlap, which is proportional to the unperturbed on-site energy of $\sim 10^4\,{\rm cm}^{-1}$. Therefore, the effects of pigment-protein interaction were neglected when obtaining the results presented in the main text. To verify whether neglecting of on-site disorder due to protein interactions is justifiable, we have also calculated the optimal size for $N=7$ pigments with random on-site disorder added to each random sample of chromophores. The results (see \ref{app:protein_interaction}) show that disorder of such magnitudes indeed has no gross effect on the results.

Each chromophore in our model is represented as a disc -- a thin cylinder -- of radius $r$ and height $a$. Each disc is supposed to represent an approximate size of the electronic cloud of the orbitals involved in the EET (highest occupied electronic orbital, lowest unoccupied electronic orbital). We have estimated the height to be $a=1\,\AA$, while the cylinder radius is taken as $r=4\,\AA$. Size of this cylinder in comparison to a BChl pigment molecule can be seen in figure~\ref{fig:ds_skica}a. Radius $r=4\,\AA$ is chosen so that it contains 16 closest non-hydrogen atoms to the ${\rm Mg}$ atom located in the center of the pyridine ring. For given locations and orientations of discs, we then determine if there are any overlaps between discs. If two discs overlap, for an example see figure~\ref{fig:ds_skica}b, we rescale the radius of one of them to the new radius $r_n$ so that they do not overlap anymore but instead only touch. After eliminating all overlaps we end up with disc's radii $r_n$, for an example see figure~\ref{fig:ds_skica}c. 

Provided the radius of the $n$-th disc $r_n$ is different from the non-overlapping size $r$, we have to appropriately rescale the on-site energy $\epsilon_n$. If the effective size of the electronic cloud is reduced from $r$ to $r_n$, the kinetic energy of electron increases by a factor $r^2/r_n^2$. We therefore estimate that the energy of excitation on a resized disc will also scale quadratically with its size, giving the on-site energy dependence
\begin{equation}
  \label{eq:1}
  \frac{\epsilon_{n}}{\epsilon^{(0)}}=\left(\frac{r}{r_n}\right)^2 - 1
\end{equation}
where $\epsilon^{(0)}$ is the excitation energy of non-deformed pigments, i.e., the energy difference of two lowest electronic states on a pigment. In FMO $\epsilon^{(0)}$ is approximately $\epsilon^{(0)} \approx 12\,300\,{\rm cm}^{-1}$. The overall offset of on-site energies is irrelevant for the dynamics in the model, therefore we shift all energies by $\epsilon^{(0)}$. Such quadratic on-site energy scaling can be rigorously shown under an assumption that the electronic eigenfunctions of the rescaled pigment are just the rescaled eigenfunctions of the original pigment of radius $r$. Let orbitals $\psi_i$ be the eigenfunctions of the Hamiltonian $H(\bm x) = T(\bm x) + U(\bm x)$, where $T(\bm x)$ is kinetic energy operator and $U(\bm x)$ is a confining potential. The on-site energy of a given chromophore $\epsilon_n$ is the difference between the energy of ground and 1st excited state, $\epsilon_n = E_2-E_1$. Assuming that eigenstates $\psi_i$ are just scaled to a smaller volume, $\psi^*_i(\bm x) = \lambda^{3/2}\psi_i(\bm x / \lambda)$, the scaling of on-site energies from equation \eref{eq:1} is obtained by comparison of eigenvalue equations for the original eigenstate $H \psi_i = E_i \psi_i$ and the scaled eigenstate $H^* \psi_i^* = E_i^* \psi_i^*$, where the scaled confining potential in $H^*$ has to be $U^*(\bm x) = U(\bm x/\lambda)/\lambda^2$.

Dipole strength of the chromophores is similarly scaled linearly with the radius $r_n$ of the cylinder,
\begin{equation}
  \label{eq:2}
  \bm d_n = \frac{r_n}{r} \bm d,
\end{equation}
where $\bm d$ is the bare transition dipole moment of the original chromophore of size $r$, and $\bm d_n$ is the scaled dipole strength of the resized disc. This can be justified on the same grounds as the scaling of on-site energies, by inserting the rescaled wavefunction $\psi^*_i(\bm x)$ into the expression for transition dipole matrix of relevant chromophore transition, $\bm d = \langle \psi_g| e \bm x |\psi_e \rangle$.

There are different possibilities of how to precisely resize discs in order to avoid overlaps. While  different procedures lead to different on-site energies, the determined optimal complex size changes by little. Results in the main text were obtained by sequentially inspecting each pair of discs, resizing only the disc having greater radius afterwards, while keeping the other disc intact. We have verified other resizing procedures, for instance, resizing both discs in pair to the same size. Such resizing effectively reduces disorder of on-site energies as even strongly overlapping pigments will have identical on-site energies. Nevertheless, for such resizing procedure, the determined optimal radius of PPCs are within $2 \, \AA$ of the values obtained by the resizing procedure used throughout the paper, and are thus within the error bounds of the model.

To summarize, in our overlapping disc model the matrix elements of $H$ are calculated for given PPC configuration (positions, as well as disc and dipole orientations) by first resizing all overlapping discs, obtaining rescaled radii $r_n$ and then scaling dipole strengths and on-site energies according to equations (\ref{eq:1}) and (\ref{eq:2}).

\subsection{Optimality criteria}

We have already introduced equations of motion that govern the dynamics of excitation on chromophores, as well the overlapping disc model that allows us to determine the Hamiltonian for a given configuration of chromophores. What is left are criteria that will enable us to determine whether a given configuration of chromophores is \textit{efficient} in terms of EET. The efficiency of the PPC complex is characterized by the probability that the excitation, initially localized on the input site, will be funneled to the reaction center trough the output site. For an example of time evolution see \ref{app:time_evolution}. The efficiency in the model is not unity because the excitation can be lost. The probability that the excitation will be transported to the reaction center can be expressed as
\begin{equation}
  \label{eq:efficiency}
  \eta = 2 \kappa \int_0^\infty {\rm d}t \, \rho_{NN} (t),
\end{equation}
which will be used as our main efficiency criterion. Closely related is the \textit{average transfer time}, which signifies the speed of transfer of excitation to the reaction center, and is expressed as
\begin{equation}
  \label{eq:att}
  \tau = \frac{2 \kappa}{\eta} \int_0^\infty {\rm d}t\, t\, \rho_{NN} (t),
\end{equation}
with smaller transfer times being better.

As an additional viability criterion of PPC, robustness of efficiency to static disorder will be also inspected. Dynamic disorder due to thermal motion is already effectively described by the dephasing Lindblad terms in equation~\eref{eq:5}. Static disorder due to structural changes of PPC, for instance due to changes in biological environment, like temperature, electric charges, etc., should be treated separately. A given configuration of pigments in PPC is \textit{robust to the static disorder} if random displacements of pigments from the original locations do not induce large changes in PPC's efficiency $\eta$ (or equivalently, the average transfer time $\tau$). To put it on a more quantitative ground, we define the \textit{pigment configuration robustness} $\sigma_\eta(\bm x)$ for a given configuration of pigments\footnote{We omitted the disc and dipole orientations from the definition of pigment configuration robustness to simplify the expressions. However, no qualitative differences are to be expected if orientations are also varied when inspecting the robustness.} with positions $\bm x=(\bm x_1, \bm x_2,..., \bm x_N )$, as a standard deviation of efficiency $\eta$ when pigment coordinates are varied in the neighborhood of original positions,
\begin{equation}
  \label{eq:9}
  \sigma_\eta^2(\bm x)=\int ( \eta^2 (\bm x+\bm y) - \bar \eta(\bm x)^2 ) w(\bm y) \, d \bm y,\quad \bar \eta(\bm x)=\int \eta(\bm x + \bm y) w(\bm y) \, d \bm y,
\end{equation}
Probability density $w(\bm y)$ defines the \textit{neighborhood} of a given configuration, and is localized around the original location of the pigments. The most straight-forward choice for the distribution $w(\bm y)$ is a product of uncorrelated normal distributions at each pigment location,
\begin{equation}
  \label{eq:10}
  w(\bm y) = (2 \pi \sigma^2)^{-N/2} \exp\left (\frac{ \bm y \cdotp \bm y }{2 \sigma^2} \right),
\end{equation}
where $\sigma$ defines the size of neighborhood in which the robustness is being probed. With given probability distribution, the robustness $\sigma_\eta$ is a function of original pigment locations $\bm x$ and size of deviations from original locations $\sigma$. In the limit of small deviations, $\sigma \rightarrow 0$, the expression can be simplified to
\begin{equation}
  \label{eq:11}
  \sigma_\eta^2 (\bm x) = \sigma^2 \sum_{i=1}^{3N} \eta_i(x_i)^2, \quad \eta_i(x_i)=\pd{\eta}{x'_i}\Big|_{x'_i=x_i},
\end{equation}
where the sum goes over all components of pigment coordinates.
The robustness $\sigma_\eta$ in the limit of small pigment displacements is thus proportional to the amplitude of pigment displacements.

\section{Optimal number of pigments: the case of Fenna-Matthews-Olson complex}

\begin{figure}[tb]
  \includegraphics{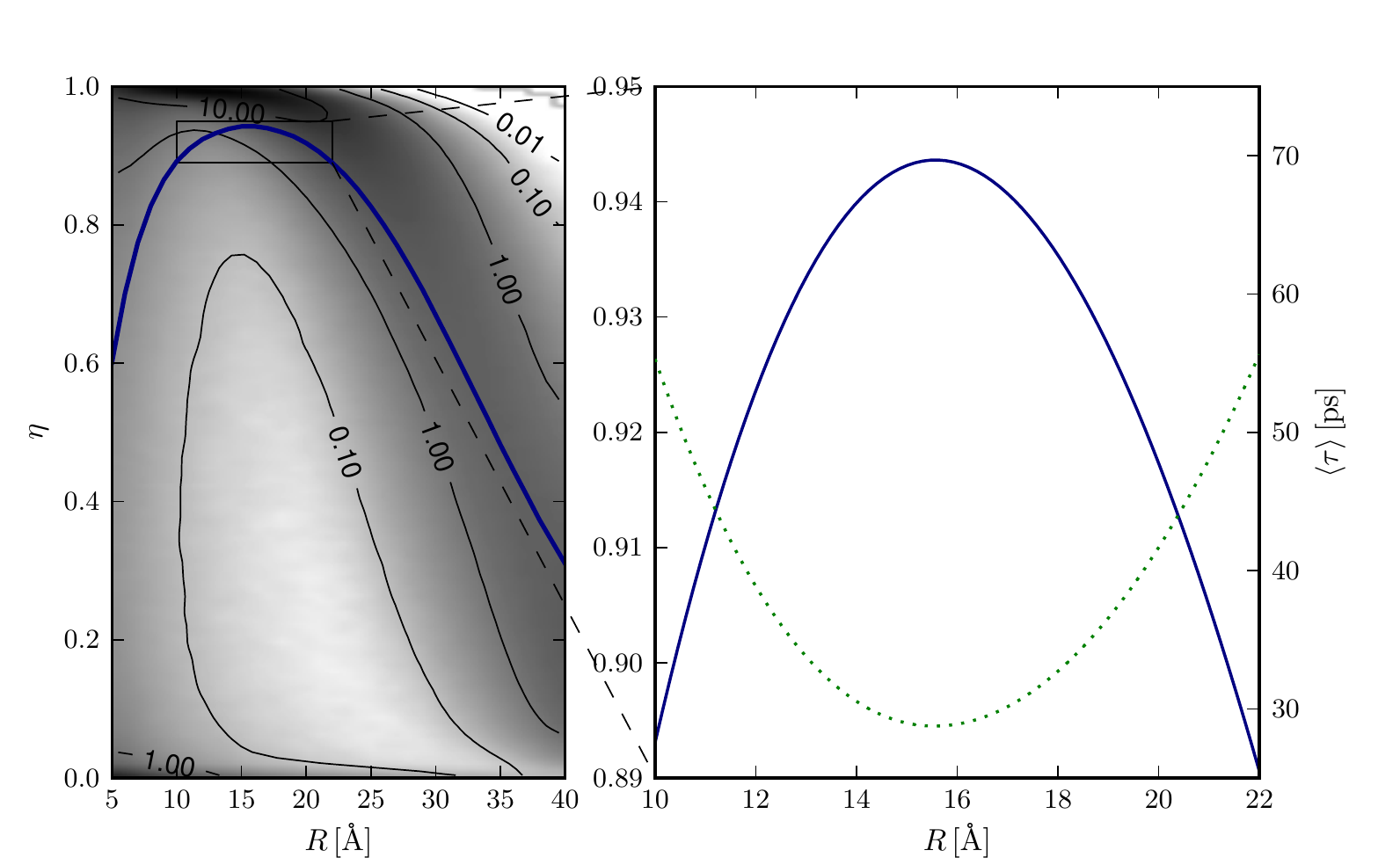}
  \caption{Probability density $p_{R, N}(\eta)$ for $N=7$ for a range of sphere radii $R$ (density plot in the background of the left plot; contours connecting equal values of $p_{R,N}$ are also shown). Solid blue curve is the average efficiency $\braket{\eta(R)}$. On the right a close-up of $\braket{\eta}$ and the average transfer time $\braket{\tau}$ (dotted curve, right axis) around the maximum of $\braket{\eta}$ is plotted.}
  \label{fig:eta_hist}
\end{figure}

\begin{figure}[tb]
  \includegraphics{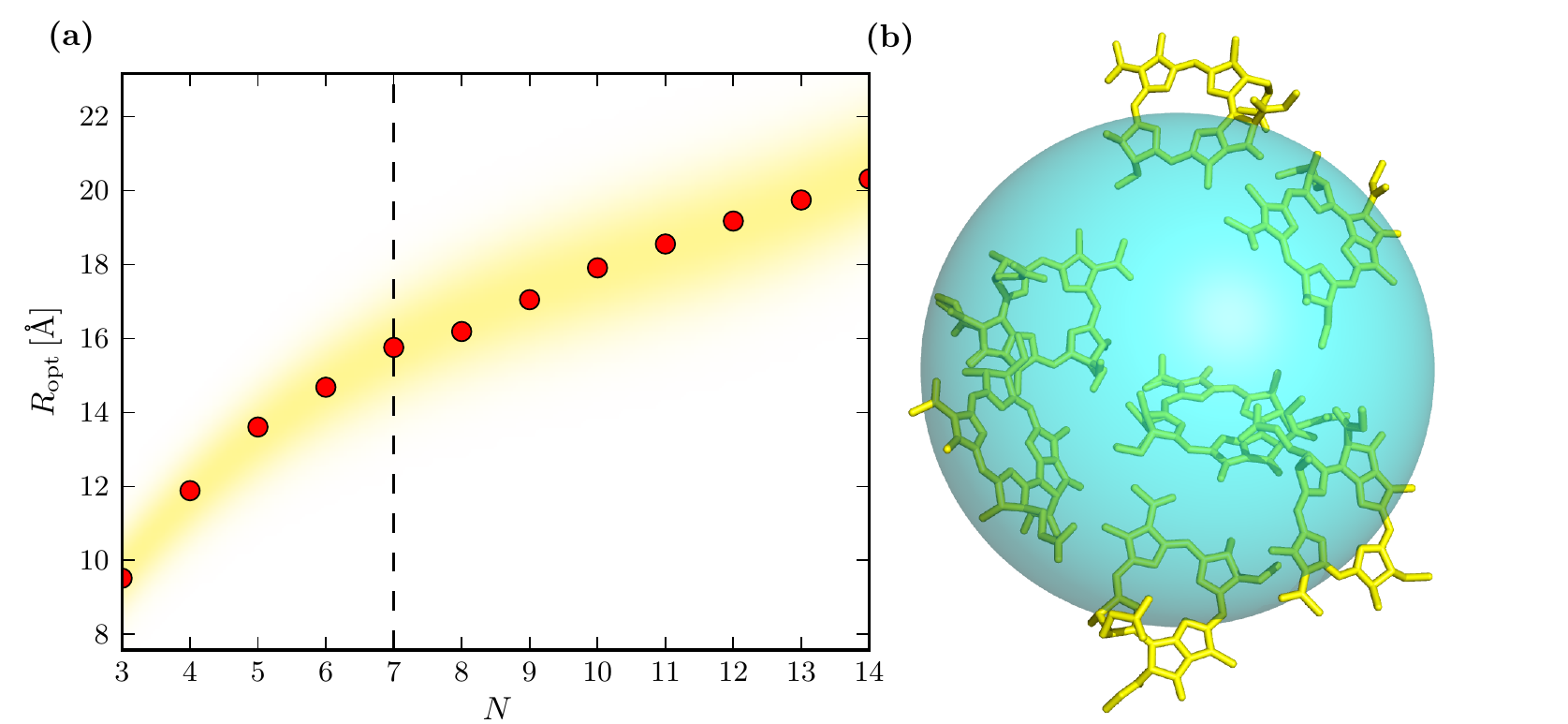}
  \caption{{\bf (a)} The optimal enclosing radius $R_{\rm{opt}}(N)$ of the sphere for different number of chromophores $N$, based on the maximum of the average efficiency $\braket{\eta(R)}$. The width of the shading (i.e. $\approx \pm 1 \AA$) denotes a range of $R$ for which the average efficiency is within 1\% of maximal value. Vertical dashed line marks the case of FMO with $N=7$ chromophores. {\bf (b)} Structure of FMO complex as determined from spectroscopic data, enclosed into the optimal sphere of the radius $R \approx 16\, {\rm \AA}$. The structural data was obtained from PDB entry 3EOJ \cite{Tronrud2009} }
  \label{fig:fmo_in_sphere}
  \label{fig:rs_max_vs_n}
\end{figure}

In this section, the efficiency and robustness of random configurations within the model will be considered on an example of FMO complex. FMO consists of $N=7$ BChl pigments. On-site energy for BChl was chosen to be $\epsilon^{(0)}=12\,300 \, \rm{cm^{-1}}$, which is within the range of on-site energies for BChls in FMO as determined in the literature \cite{Adolphs2006,SchmidtamBusch2011, Olbrich2011}. The strength of transition dipole moment $d=|{\bm d}|$ was taken as $d^2 = 26 \, \rm{D^2}$ \footnote{Dipole moment is given in units of debye ($\rm D$). Conversion of interaction $V_{mn}$ to units of $\rm cm^{-1}$ is obtained by conversion $\frac{{\rm D}^2}{4 \pi \varepsilon_0} = 5030 \, {\rm cm^{-1}\AA^3}$.} (note that published values for $d$ from calculations and experimental data vary considerably \cite{Milder2010}). On-site energy $\epsilon^{(0)}$ and dipole strength $d$ used hold for BChl pigments in general and therefore the results presented are expected to be valid also for other PPCs containing BChls, not just for the FMO complex.

To determine the optimal size of PPC for a given number of chromophores (or equivalently, optimal number of chromophores for a given size), we considered two criteria based on overall behavior of efficiencies and robustness of random configurations. In the following, the motivation for choosing such optimality criteria will be given, and the results for the case of FMO will be presented.

\subsection{Average efficiency}

We shall use the average efficiency $\braket{\eta}$, averaged over random positions and orientations of pigments enclosed in a predefined volume. The reason to use random averaging with uniform distribution is twofold: first, high average EET efficiency under uniform averaging will mean that there are many different configurations that have high efficiency, i.e., high efficiency is globally robust. Choosing averaging over random configurations therefore offers insight in how special efficient configurations of chromophores are within the space of all configurations. Second reason is that we have a priori no knowledge what would be the appropriate measure for possible pigment configurations under say different protein cage foldings due to for instance mutations. A uniform measure represents in this case a ``least-information'' distribution. \footnote{We have to note that we also checked other distributions, for instance a uniform distribution on the surface of a sphere of radius $R$, and obtained practically the same results. For instance, the difference in the position of the maximum in figure~\ref{fig:eta_hist} was within our error estimate of $1\,\AA$ (seen as an ``error'' band in figure~\ref{fig:rs_max_vs_n}).} Using configurations sampled according to a uniform distribution over chromophore positions within a ball of radius $R$ and random orientations of dipoles and discs, the \textit{average efficiency} is calculated. Formally, it can be written as
\begin{equation}
  \label{eq:12}
  \braket{\eta(R)} =\int_{R} \eta (\bm X) w_{\rm{conf}}(\bm X) \, d \bm X,
\end{equation} where $\bm X$ contains positions and orientations of chromophore discs and dipoles (apart from the positions of input and output sites which are fixed on the poles of the sphere), and $w_{\rm{conf}}(\bm X) \propto 1 $ signifies a uniform distribution of chromophores inside the sphere. Observing the dependence of the average efficiency $\braket{\eta(R)}$ for different number of chromophores and different radii $R$ of the enclosing sphere, we can determine the optimal number of chromophores for a given radius $R$, or equivalently, the \textit{optimal radius} $R_{\rm opt}$ for a given number of chromophores.

To obtain a more detailed information about efficiencies of random configuration, we also observed the probability distribution over efficiencies $p_{R, N}(\eta)$, defined as $p_{R,N}(\eta)=\int_R \delta(\eta(\bm X)-\eta) w_{\rm conf}(\bm X) d \bm X.$
For the number of chromophores as found in FMO ($N=7$), the probability distribution over efficiencies $p_{R,7}(\eta)$ is shown in figure~\ref{fig:eta_hist}. When going from large radii $R$ to smaller, configurations tend to get more efficient, which is expected as the chromophores are closer to each other, thus increasing the dipole coupling. However, as $R$ is reduced even further, overlapping of chromophores gets more probable, causing an on-site energy disorder. This leads to the localization of excitation on chromophores not connected to the sink site. Such configurations have low efficiency. Therefore, as $R$ gets smaller the distribution $p_{R,N}$ becomes bimodal, with lower efficiency mode due to overlapping configurations and high efficiency mode for non-overlapping configurations.

Low efficiency of overlapping configurations therefore leads to a maximum in the average transfer efficiency $\braket{\eta(R)}$ at the optimal radius $R_{\rm opt}$. The average transfer efficiencies at the optimal radius are rather high, e.g. for $N=7$ in figure \ref{fig:eta_hist} it is $\braket{\eta(R_{\rm{opt}})} \approx 0.95$ with a large fraction of configurations having even larger efficiency than the average. Thus within the model, high efficiency is not due to finely tuned pigment positions and orientations, but occurs for majority of pigment configurations for parameters estimated to be relevant in PPCs. For the FMO case with $N=7$, the optimal radius was estimated to $R_{\rm{opt}}\approx 16 \, \rm{\AA}$, which fits the actual configuration of pigments very well (see figure~\ref{fig:fmo_in_sphere}). The average transfer time $\braket{\tau}$ is also minimal at $R = R_{\rm opt}$ (see figure \ref{fig:eta_hist}). Optimal average transfer time of $\sim 30\,{\rm ps}$ is so large due to the contribution of very inefficient configurations of chromophores. Looking at the average transfer time of the $5 \%$ of most efficient configurations, we get the value of $~5\,{\rm ps}$, which is comparable to the transfer times as determined using different models of the FMO in references \cite{Rebentrost2009,Mohseni2008,Caruso2009,Kreisbeck2011}.

The estimated optimal radius is quite insensitive to small variations of input parameters, e.g. dipole moment $d$, chromophore disc radius $r$ or its thickness $a$, or the scaling of on-site energies and dipole strengths of resized discs. For instance, decreasing disc radius to $r=3.5\,\AA$ decreases $R_{\rm opt}$ by $\approx 2 \, \AA$, changing disc thickness to $a = 0.5\,\AA$ or $1.5\,\AA$ changes $R_{\rm opt}$ by $\approx \mp 2\, \AA$, while changing quadratic energy scaling to a linear or cubic one again changes $R_{\rm opt}$ by $ \approx \mp 2 \, \AA$. Similarly, changing the dephasing rate $\gamma$ by a factor of 2 changes $R_{\rm opt}$ by $\approx 2 \, \AA$, see \ref{app:gamma}. Details of the disc resizing procedure also change $R_{\rm opt}$ for less than $2\,\AA$, as the extreme case of resizing each overlapping disc pair to the same size reduces $R_{\rm opt}$ by $\approx -2\,\AA$.

The optimal radius of the enclosing sphere was obtained from the average efficiency over \textit{all} random configurations within a sphere. However, even though the evolutionary drive to more efficient configurations might not be very strong if \textit{majority} of configurations are already efficient, still some optimization is to be expected. Thus one might argue that the \textit{optimal} enclosing volume of the natural PPCs should be determined considering only the ensemble of \textit{more} optimal configurations. We will denote such averages with $\braket{\eta}_p$ where $p$ specifies a portion of most efficient configurations that should be taken into account when calculating the average (e.g. $\braket{\eta}_{0.05}$ is the average of $\eta$ over 5\% of most efficient configurations as shown in figure \ref{fig:stds_etas_vs_sizes}). As $p$ is reduced, the overall value of average efficiency $\braket{\eta}_p$ will increase. The increase will be more pronounced in the region of $R < R_{\rm{opt}}$, where the distribution is bimodal. The location of the maximum of $\braket{\eta}_p$ will be thus moved to smaller values of $R$, indicating more densely packed chromophore configurations. However, as we will see in next subsection, robustness of such densely packed configurations deteriorates very quickly, supporting our choice of estimator for the $R_{\rm opt}$.

Note that the overlaps between pigments and protein cage are not considered explicitly in the model. If overlaps with protein cage would be taken into account, $R_{\rm opt}$ would represent the size of a protein cage, whereas in our model without pigment-protein overlaps, $R_{\rm opt}$ is the size of a sphere that contains all pigment centers. For instance, looking at figure~\ref{fig:fmo_in_sphere}b, we can see that the sphere with $R_{\rm opt}$ contains all pigment centers, while parts of few pigments still protrude the bounding sphere. If overlaps of pigments with the protein cage would be taken into account explicitly, $R_{\rm opt}$ would be approximately by a disc radius $r=4\,\AA$ larger, i.e. in corresponding figure, the bounding sphere would enclose all pigments completely.

\subsection{Robustness}

\begin{figure}[bt]
  \includegraphics{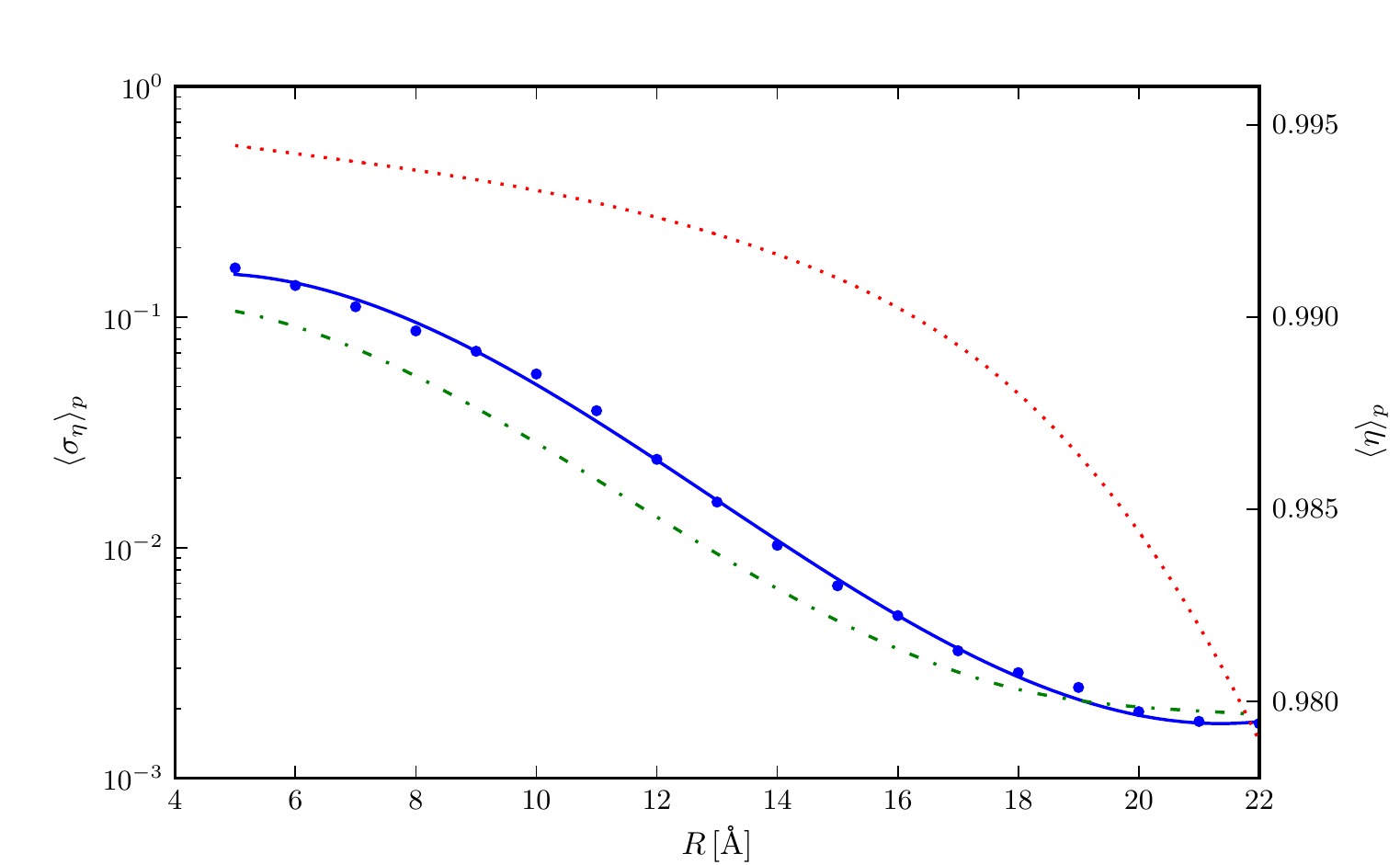}
  \caption{The average robustness of top 5\% of most efficient configurations $\braket{\sigma_\eta}_{0.05}$ (solid blue curve) as a function of the enclosing sphere radius $R$. Dash-dotted green curve represents the average robustness of the top 15 \% of most efficient configurations $\braket{\sigma_\eta}_{0.15}$. Dotted red curve is the average efficiency over the top 5\% of configurations $\braket{\eta}_{0.05}$ (right axis).}
  \label{fig:stds_etas_vs_sizes}
\end{figure}

Robustness of PPC configurations to static disorder should also be taken into account when determining whether a given configuration of pigments is feasible, as the conditions in which PPCs operate are subject to constant environmental changes. In previous subsection, we have inspected the probability distribution of efficiencies $\eta$ over random configurations, showing that majority of random configurations achieve relatively high efficiency when the enclosing volume is \textit{optimal}. In this subsection, we will present analogous analysis of the robustness of random configurations, in particular of those with high $\eta$. We shall show that highly efficient configurations in small enclosing $R$ are very fragile.

We have defined robustness of efficiency $\sigma_{\eta}$ in equation~\eref{eq:9}. In simulations we have displaced the pigment positions according to normal distribution with a width of $\sigma=0.1 \, \rm{\AA}$, which is small enough to quantify the robustness in the neighborhood of specific configuration, while larger than displacements due to thermal vibrations, which are already effectively described by the Lindblad equation. We are specifically focusing on a subset of the \textit{most} optimal configurations in terms of $\eta$. The \textit{average} robustness of the subset of optimal configuration is denoted as $\braket{\sigma_\eta}_p$, where $p$ specifies a fraction of most optimal configurations in terms of EET efficiency $\eta$. As an example, we will consider robustness of top 5\% of efficient configurations $\braket{\sigma_\eta}_{0.05}$. The dependence of the average robustness on the radius of the enclosing sphere $R$ is shown in figure~\ref{fig:stds_etas_vs_sizes}. The average efficiency of optimal configurations $\braket{\eta}_{0.05}$ is also shown in the figure. While the average efficiency of top 5\% of optimal configurations continues to rise as the enclosing sphere radius $R$ is reduced, we can see that the average robustness $\braket{\sigma_\eta}$ worsens very quickly as the $R$ drops below $R_{\rm{opt}}$.

Quick worsening of EET robustness with reducing sphere radius suggests that even if the PPC configurations occurring in nature are indeed optimized in terms of pigment positions and orientations, the excessive stacking of pigments is not favored as it makes PPC configurations very sensitive to any displacements of pigments. The transition from robust to non-robust regime takes place at a radius comparable to $R_{\rm{opt}}$ at which the average efficiency $\braket{\eta}$ has a maximum. This is not surprising as both, the efficiency of configurations and robustness of configurations, are strongly influenced by the overlapping of pigments which gets more pronounced for $R \lesssim R_{\rm{opt}}$.

We have presented results for the robustness $\sigma_\eta$ of the 5\% of most efficient configurations, with pigment displacements $\sigma=0.1 \, \rm{\AA}$. General characteristics of $\braket{\sigma_\eta}_p$ however do not quantitatively change for different $p$ (the case of $p=0.15$ is also shown in figure~\ref{fig:stds_etas_vs_sizes}) or displacements $\sigma$. Most importantly, the radius $R$ at which the robustness of configurations drops significantly takes place at approximately the value of $R_{\rm opt}$. Same behavior of robustness is observed also in the limit of \textit{infinitesimal} robustness from equation~\eref{eq:11} where $\sigma \rightarrow 0$.

\section{Optimal pigment numbers in other PPCs}
\label{sec:comparison}

In previous section we have calculated the optimal size $R$ or the optimal number of pigments for the FMO complex. We also demonstrated that a large portion of chromophore configurations has high efficiency when the enclosing volume is properly chosen ($\sim R_{\rm{opt}}$). Additionally, robustness of configurations to chromophore displacements starts to deteriorate quickly once the enclosing volume is reduced below $R_{\rm opt}$. Based on these two observations, we argue that the enclosing volume of PPCs occurring in nature should be close to the optimal volume as determined by our simple model. In this section we will present similar results for the PPCs containing chlorophyll (Chl) chromophores.

We compare results of the model to the structure of PPCs from the Photosystem II (PSII) \cite{Croce2011}, found in cyanobacteria, algae and plants. PSII consists of multiple functional units, which are either part of the outer light-harvesting antenna or the inner core, to which excitations are funneled. In the light-harvesting antenna we will consider the light harvesting complex II (LHCII), while in the core we will focus on the PC43 and PC47 complexes that funnel excitations to the reaction center and thus have a similar role as the FMO complex in bacteria. A monomeric unit of LHCII contains 14 chlorophyll molecules (8 Chl-\textit{a} and 6 Chl-\textit{b}), while CP43 and CP47 contain 13 and 16 Chls respectively.

Model parameters for the sink rate, dephasing and recombination are kept the same as in the FMO case, while the transition dipole strength and on-site energies are different for Chl molecules. Transition dipole moment of Chl molecules is chosen as $d^2 = 15 \, \rm{D^2}$ and the on-site energy $\epsilon^{(0)}=15\,300\, \rm{cm^{-1}}$, where values were taken according to reference~\cite{Novoderezhkin2011} (we take the average between values for Chl-\textit{a} and Chl-\textit{b}). For the CP43 and CP47 complexes we have simulated random configurations of 13 and 16 chromophores enclosed into a sphere as the actual chromophore positions are distributed relatively uniformly in all directions. The shape of LHCII is however significantly elongated in one direction. We therefore choose the cylindrical volume, having only one additional parameter that has to be provided, i.e. the ratio between the cylinder radius $R_c$ and cylinder height $A$. Based on positions of the LHCII chromophores we have estimated the ratio of the two to be $R_c/A = 0.34$.

\begin{figure}[ht]
  \includegraphics{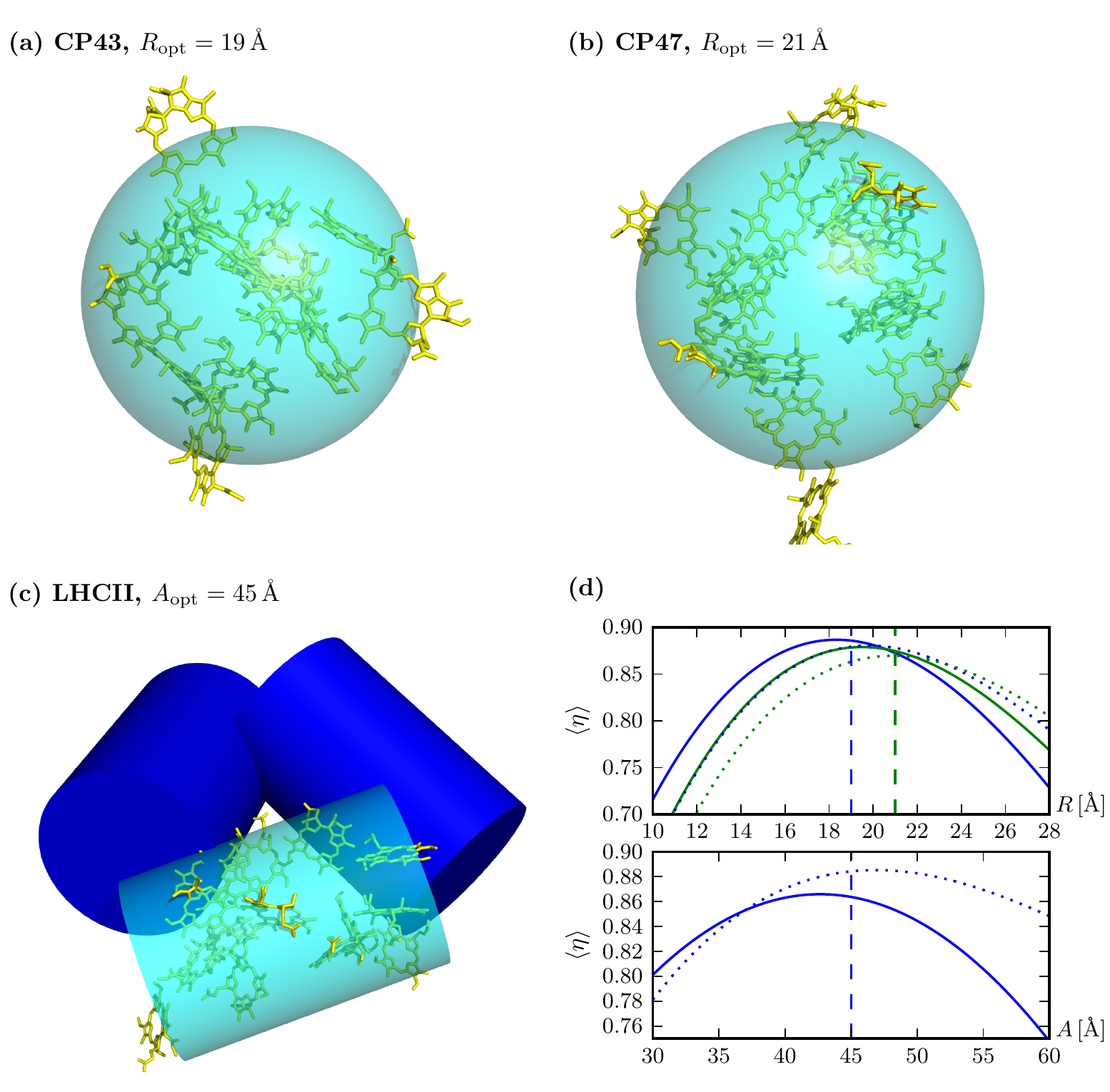}
  \label{fig:ppcs_in_geometries}
  \caption{{\bf (a) - (c)} Various pigment-protein complexes (PPCs) enclosed in optimal geometries as described in the main text. The structural data was obtained from PDB entries 1RWT (LHCII) \cite{Liu2004} and 3ARC (CP43, CP47) \cite{Umena2011}. In (c) enclosing geometries of two additional monomeric units of the LHCII trimer are also shown. {\bf (d)} Plot showing dependence of the average efficiency $\braket{\eta}$ on the size of the enclosing geometry. The upper panel shows the case of CP43 (green) and CP47 (blue), while the lower panel shows the case of LHCII. Solid curves are for the fixed input site, while dotted line for the randomly placed input site. Vertical dashed lines mark the sizes of enclosing volumes as used in subfigures (a) - (c). }
\end{figure}

The CP43 and CP47 primarily play a role of an exciton wire, making the model with input site and output site at the opposite sides of the sphere applicable. The optimal radius as predicted by the model is $R \approx 18 \, \rm{\AA}$ for CP43 and $R \approx 20 \, \rm{\AA}$ for CP47. As the LHCII also has to transport excitations from adjacent complexes, we have also determined the optimal shape of LHCII with input and output sites located at the opposite sides of the enclosing cylinder. With the ratio $R_c/A$ fixed, we have varied the height $A$ of the cylinder and determined the \textit{optimal} height to be $A_{\rm{opt}} \approx  43 \, \rm{\AA}$.

In addition to acting as an excitation wire, CP43 and CP47 complexes are also directly involved in the absorption of photons, in which case the role of the input site can be taken by any chromophore site. This is even more common scenario in the LHCII complex, whose primary role is the absorption of photons. To verify whether the findings about optimal enclosing volume are also valid when the main purpose of PPC is the absorption of photons, we randomly placed the input site inside the enclosing geometry for each configuration in random ensemble. General characteristics of the distribution over efficiencies $p_{R}(\eta)$ do not change considerably, however, the distribution is somewhat shifted to the higher efficiencies because in many random configurations the input site is considerably closer to the output site than the diameter of the enclosing volume. This results in the optimal size of enclosing volume being somewhat larger, $R_{\rm{opt}} \approx 20 \rm{\AA}$ for CP43 and $R_{\rm{opt}} \approx 22 \rm{\AA}$ for CP47. For the LHCII we have moved the output site to the midpoint on the side between top and bottom of the cylinder, where the actual output site is supposedly located \cite{Renger2011}. For such geometry and previously used $R_c/A = 0.34$, we have obtained the optimal height of the enclosing cylinder at $A \approx 47 \rm{\AA}$.

The optimal enclosing values obtained from the model (averaged between the case for fixed input site and random input site) were compared to the actual configurations of pigments as obtained from spectroscopic data, and are shown in figure~\ref{fig:ppcs_in_geometries}a-c, showing good agreement. For the spherical geometries, we have centered the sphere of optimal radius $R_{\rm{opt}}$ to the arithmetic mean of locations of BChl/Chl centers. For the LHCII, where cylinder was used as the enclosing geometry, the cylinder axis was determined such that $\sum_i^N r_{\perp i}^2$ was minimal, where $r_{\perp i}$ is the distance from the cylinder axis to the position of $i$-th Chl center. Interestingly, three cylinder axes do not lie in a plane but are instead tilted at an angle $15 ^\circ$ to the plane containing three cylinder centers. It is not known if this plays any functional role.

\section{Conclusion}

We have studied the efficiency of excitation energy transfer in protein-pigment complexes for random configurations of pigments. The Hamiltonian part of Lindblad master equation is determined from the geometry of pigment configurations. If pigments are too close, so that they overlap, this introduces a disorder in on-site energies, effectively inhibiting excitation transport. Fixing the enclosing volume in which pigment molecules are located we have calculated the average efficiency over random pigment configurations as well as robustness of efficiency to variations of pigment locations. Doing this we have determined the optimal number of pigments for a given size of the complex. Even though the model is an oversimplification of actual processes that take place in nature, statistical predictions obtained from the model are robust to its variations.

Comparing theoretically predicted optimal number of pigments with several naturally-occurring complexes we find good agreement. This might indicate that PPCs are not optimized just to have the highest possible efficiency -- in fact, efficient configurations are quite common -- but instead to be robust to variations in pigment locations. Namely, it turns out that configurations optimized for the highest efficiency, that is those with specially tuned positions and dipole orientations, are very sensitive to small perturbations. The number of pigments in nature is therefore chosen in such a way that the probability of having efficient configurations that are at the same time also robust is the highest.

The presented findings could be in principle verified experimentally by modifying the structure of known PPCs and probing the efficiency of excitation transfer. For the FMO complex the structure was already changed by mutation of genes encoding the structure of BChls, as well as by substituting the carbon $^{12}\rm C$ atoms with $^{13}\rm C$ \cite{Hayes2011}. Comparison of excitonic spectra revealed no distinctive differences in the dynamics of excitations, complying with the hypothesis that configurations are not highly tuned but instead very robust. An additional intriguing possibility would be also to inspect the characteristics of FMO with mutated protein cage, modifying positions and orientations of pigment molecules. One could also compare our predictions for the optimal sizes (e.g. figure~\ref{fig:rs_max_vs_n}a) with other complexes occurring in nature.

\appendix
\section{Dependence on the dephasing rate $\gamma$}
\label{app:gamma}
\setcounter{section}{1} 

\begin{figure}[bt]
  \includegraphics{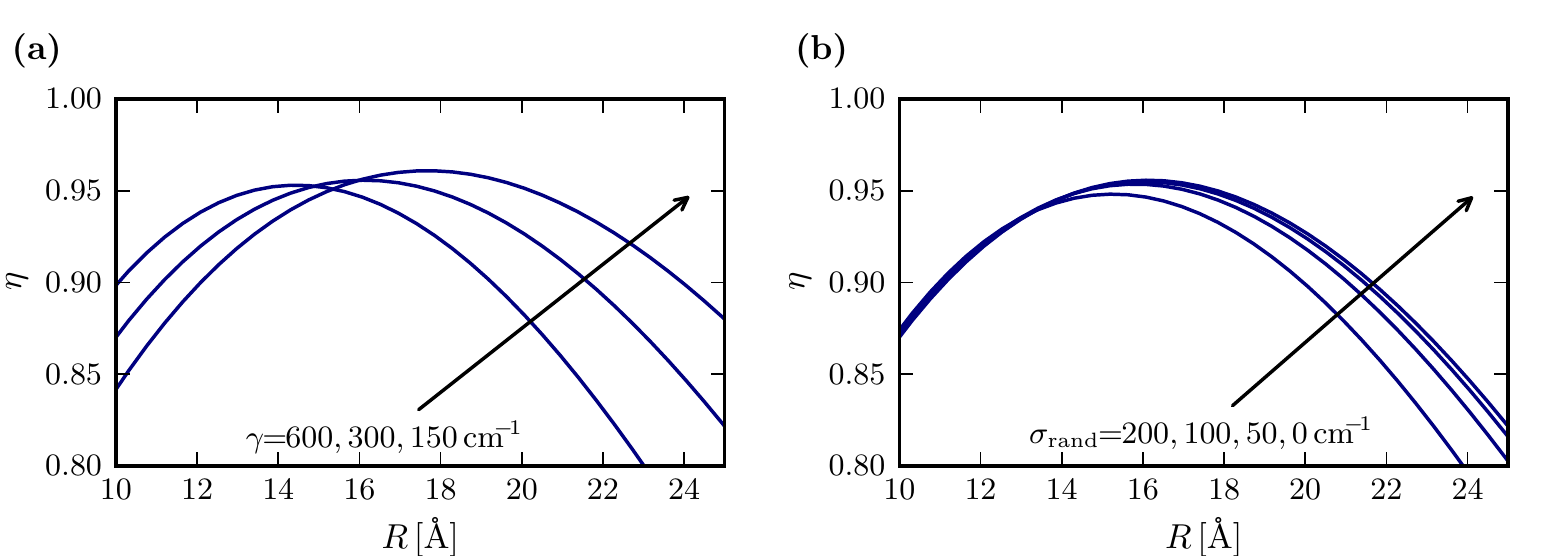}
  \caption{{\bf (a)} Average efficiencies $\braket{\eta}$ for different values of dephasing $\gamma = 150\,{\rm cm}^{-1},\, 300\,{\rm cm}^{-1},\, 600\,{\rm cm}^{-1}$. {\bf (b)} Average efficiencies $\braket{\eta}$ for different values of random on-site energies disorder $\sigma_{\rm rand}$. }
  \label{fig:appendix_fig}
\end{figure}

In the simplified model used, environmental interaction is described by dephasing rate $\gamma$, having the same value for all chromophore sites. In principle, environmental interaction requires more involved equations of motion (e.g. HEOM \cite{Ishizaki2009}), taking into account the spectral density of the environmental modes. However, due to crude nature of the model, simplified description is expected to account for main environmental effects influencing the efficiency of exitation transfer (see e.g. \cite{Wu2010, Wu2011} for more detailed comparison of approaches). The adequacy of simple Lindblad-type description of dynamics for determination of optimal size is also justified due to the high robustness of the results to the actual choice of dephasing value $\gamma$, as seen in figure~\ref{fig:appendix_fig}a, where $R_{\rm opt}$ only changes for $\sim \pm 2 \AA$ as dephasing rate $\gamma$ is changed by a factor of 2. Optimal size $R_{\rm opt}$ of the PPC is somewhat smaller as dephasing rate gets stronger, which is expected as larger dephasing rate enables transfer across the sites with greater on-site energy mismatch, getting increasingly common in more compact configurations of chromophores.

\section{Random on-site disorder}
\label{app:protein_interaction}

To verify whether effects of the local chromophore environment due to e.g. pigment-protein interaction can affect the findings about the optimal PPCs sizes, we have amended the Hamiltonian in equation \eref{eq:3} with random on-site disorder $\epsilon_n^{\rm rand}$ of the magnitudes as found in naturally occurring PPCs (i.e. on-site energy differences in the order of $\sim 100\,{\rm cm}^{-1}$). The values of disorder for each realization of random PPC were calculated according to Gaussian distribution with variance $\sigma_{\rm rand}$. Results are shown in figure~\ref{fig:appendix_fig}b. In the region $R < R_{\rm opt}$, where average transfer efficiency is strongly affected by disc overlaps, an addition of random on-site energy disorder has no noticeable effect. The effect is more pronounced for $R>R_{\rm opt}$ where overlapping of discs is not the limiting factor of transfer efficiency any more. The estimated optimal size $R_{\rm opt}$ however is not changed considerably by an addition of random on-site disorder.

\section{Time evolution of site populations}
\label{app:time_evolution}

\begin{figure}[bt]
  \includegraphics{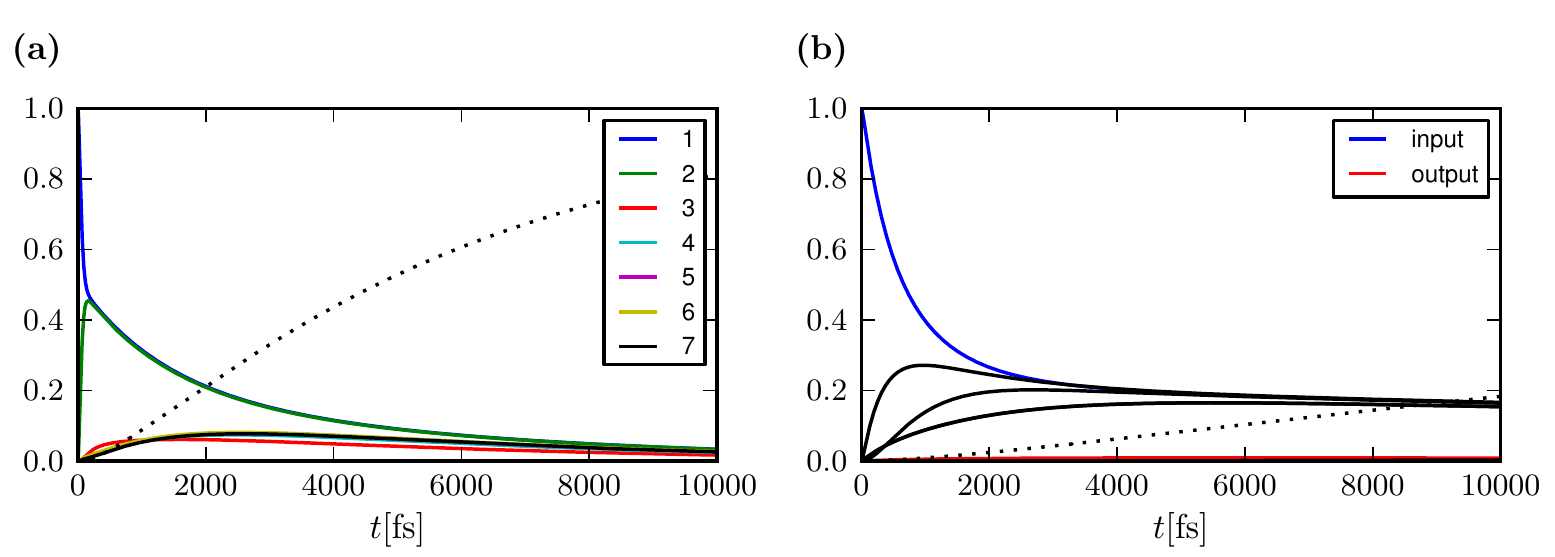}
  \caption{{\bf (a)} Time evolution of site populations $\bra{n} \rho \ket{n}$, calculated using Lindblad model for the FMO Hamiltonian from reference \cite{Cho2005}, resulting in efficiency of $\eta=0.99$ and average transfer time of $\tau=6.2\,{\rm ps}$. At $t=0$, excitation is localized on site 1. Sink is connected to the site 3. Dotted line represents the population of the sink. {\bf (b)} Time evolution for a random sample, generated for $R=18\,\AA$, with efficiency $\eta=0.90$ and average transfer time $\tau = 45\,{\rm ps}$. Blue line is the population of the input site that is initially excited. Red line is the population of the output site, connected to the sink. Dotted line represents the population of the sink. }
  \label{fig:appendix_fig_2}
\end{figure}

To provide some insight into the temporal dynamics of the excitation transport, we present the time evolution of site populations for two different realizations of PPC within the Lindblad model. In figure~\ref{fig:appendix_fig_2}a, time evolution for the FMO Hamiltonian from reference \cite{Cho2005} is shown, and in figure~\ref{fig:appendix_fig_2}b, the time evolution of randomly generated PPC of radius $R=18\,\AA$. The values of dephasing, sink rate and recombination rate are the same as used in the main text.

\section*{References}

\bibliographystyle{iopart-num}
\bibliography{references}

\end{document}